\documentclass[aps,prd,reprint,showpacs,showkeys]{revtex4-1}

\begin{document}

\title{Lorentz Violation of the Photon Sector in Field Theory Models}

\newcommand*{\PKU}{School of Physics and State Key Laboratory of Nuclear Physics and
Technology, \\Peking University, Beijing 100871,
China}\affiliation{\PKU}
\newcommand*{\CIC}{Collaborative Innovation Center of Quantum Matter, Beijing, China}\affiliation{\CIC}
\newcommand*{\CHEP}{Center for High Energy
Physics, Peking University, Beijing 100871,
China}\affiliation{\CHEP}

\author{\surname{Zhou} Lingli}
\email[email:]{ zhoull@pku.edu.cn}\affiliation{\PKU}
\author{Bo-Qiang Ma}
\email[email:]{
mabq@pku.edu.cn}\affiliation{\PKU}\affiliation{\CIC}\affiliation{\CHEP}


\date{\today}

\begin{abstract}
We compare the Lorentz violation terms of the pure photon sector
between two field theory models, namely the minimal standard model
extension (SME) and the standard model supplement (SMS). From the
requirement of the identity of the intersection for the two models,
we find that the free photon sector of the SMS can be a subset of
the photon sector of the minimal SME. We not only obtain some
relations between the SME parameters, but also get some constraints
on the SMS parameters from the SME parameters. The CPT-odd
coefficients $(k_{AF})^{\alpha}$ of the SME are predicted to be
zero. There are 15 degrees of freedom in the Lorentz violation
matrix $\Delta^{\alpha\beta}$ of free photons of the SMS related
with the same number of degrees of freedom in the tensor
coefficients $(k_F)^{\alpha\beta\mu\nu}$, which are independent from
each other in the minimal SME, but are inter-related in the
intersection of the SMS and the minimal SME. With the related
degrees of freedom, we obtain the conservative constraints
($2\sigma$) on the elements of the photon Lorentz violation matrix.
The detailed structure of the photon Lorentz violation matrix
suggests some applications to the Lorentz violation experiments for
photons.
\end{abstract}



\maketitle

\section{Introduction}

Lorentz symmetry is one of the basic principles of modern physics,
and it stands as one of the basic foundations of the standard model
of particle physics. The minimal standard model has achieved a great
success in predictions and explanations of various experiments.
Nevertheless, some fundamental questions remain to be answered. One
of the most essential questions is whether the Lorentz invariance
holds exactly or to what extent it holds. Through theoretical
researches and recent available experiments on the Lorentz
invariance violation (LIV or LV), we can obtain a deeper insight
into the nature of Lorentz symmetry and clarify these fundamental
questions.

The possible Lorentz symmetry violation (LV) effects have been
investigated for decades from various theories, motivated by the
unknown underlying theory of quantum gravity together with various
phenomenological applications~\cite{AmelinoCamelia:2008qg,ShaoMa10}. The
existence of an ``{\ae}ther" or ``vacuum" can bring the breaking
down of Lorentz invariance~\cite{Dirac,Bjorken}. From basic
consideration, there are investigations on the concepts of
space-time such as whether the space-time is discrete or
continues~\cite{p99,Snyder,Snyder1,xu-l}, or whether a fundamental length
scale should be introduced to replace the Newtonian constant
$G$~\cite{lv5}. The Lorentz violation can happen in many alternative
theories, e.g., the doubly special relativity
(DSR)~\cite{Amelino-Camelia2002,Magueijo:2001cr,Zhang2011}, torsion
in general relativity~\cite{LV-GR1,LV-GR10,LV-GR2}, non-covariant field
theories~\cite{Copenhagen,Copenhagen2,Copenhagen3,Copenhagen4}, and large
extra-dimensions~\cite{Ammosov2000,Pas2005}. Among these theoretical
investigations of Lorentz violation, it is a powerful framework to
discuss various LV effects based on traditional techniques of
effective field theory in particle physics. It starts from the
Lagrangian of the standard model, and then includes all possible
terms containing the Lorentz violation effects. The magnitudes of
these LV terms can be constrained by various experiments. The
standard model extension (SME)~\cite{SME98} is an example within
such field theory frameworks, in which the LV terms are measured
with several tensor fields as coupling constants, and
modern experiments have built severe constraints on the relevant
Lorentz violation parameters~\cite{KR08}.

The standard model extension is an effective framework for
phenomenological analysis. We still need a fundamental theory to
derive the Lorentz violation terms from basic principles. An attempt
for such a purpose has been offered in Refs.~\cite{Ma10,SMS3}, in
which a more basic principle, denoted as physical invariance or
physics independence, is proposed to extend the basic principle of
relativity.
Instead of the requirement that
the equations describing the laws of physics have the same form in
all admissible frames of reference, it requires that the equations
describing the laws of physics have the same form in all admissible
mathematical manifolds. This principle leads to the following
replacement of the ordinary partial $\partial_{\alpha}$ and the
covariant derivative $D_{\alpha}$
\begin{equation}\label{eqn:substitution}
\partial^{\alpha} \rightarrow M^{\alpha\beta}\partial_{\beta},\quad
D^{\alpha}\rightarrow M^{\alpha\beta}D_{\beta},
\end{equation}
where $M^{\alpha\beta}$ is a local matrix. We separate it to two
matrices like $M^{\alpha \beta}=g^{\alpha \beta}+\Delta^{\alpha
\beta}$, where $g^{\alpha\beta}$ is the metric tensor of space-time
and $\Delta^{\alpha \beta}$ is a new matrix which is particle-type
dependent generally. Since $g^{\alpha\beta}$ is Lorentz invariant,
$\Delta^{\alpha \beta}$ contains all the Lorentz violating degrees
of freedom from $M^{\alpha \beta}$. Then $\Delta^{\alpha \beta}$
brings new terms violating Lorentz invariance in the standard model
and is called Lorentz invariance violation matrix. This new
framework can be referred as the Standard Model Supplemented with Lorentz Violation Matrix or
Standard Model Supplement
(SMS)~\cite{Ma10,SMS3} for short, and it has been applied to discuss the
Lorentz violation effects for protons~\cite{Ma10},
photons~\cite{Ma10graal}, and neutrinos~\cite{SMS-OPERA,Ma:2011jj}.

Before accepting the SMS as a fundamental theory, one can take the
SMS as an effective framework for phenomenological applications by
confronting  various experiments to determine and/or constrain
the Lorentz violation matrix $\Delta^{\alpha \beta}$ for various
particles. From a more general sense, the SMS should be a subset of
a general version of the SME. However, in the case of the minimal
version of the SME, the relationship between the SMS and the SME is
unclear yet. The purpose of this paper is to compare the Lorentz
violation effects of the photon sector between the two models. As
have been well known, light has always played a significant role in
the developments of physics, and the Lorentz violation in the photon
sector is also under active investigations both theoretically and
experimentally~\cite{ShaoMa10}. The Lorentz violation parameters of
photons in the SME have been well constrained by various
experiments, as summarized in Ref.~\cite{KR08}. By
confronting  the collected data in Ref.~\cite{KR08}, we can
obtain bounds and detailed structure of the Lorentz violation
parameters in the model SMS~\cite{Ma10,SMS3,Ma10graal}.

This article is organized as follows. Sec.~\ref{sec_relation}
provides the relation between the photon Lorentz violation matrix in
the SMS and various tensor fields in the minimal SME, in the case
that the two models give the same results. These relations define
the boundary of the intersection of these two theories too. In
Sec.~\ref{sec_photon_lvm}, we discuss the general structure of the
photon Lorentz violation matrix in our model, and its implications
for the potential property of the space-time structure for photons.
At the same time, we obtain the constraints on the elements of the
photon Lorentz violation matrix. Then conclusion is given in the
last section.

\section{Relations between two models}\label{sec_relation}
In the standard model extension (SME), the terms that violate
Lorentz invariance are added by hands from some considerations such
as gauge invariance, Hermitean, power-counting renormalizability and
etc. We consider just the minimal SME~\cite{SME98} here. The Lorentz
violation terms in the minimal SME contain many tensor fields as
coupling constants, with their magnitudes to be
determined/constrained by experiments. Though these Lorentz
violation terms are allowed by some general considerations, the
reason for their existence still needs to be provided from
theoretical aspects. In the SMS~\cite{Ma10,SMS3}, the Lorentz
violation terms arise from the replacement of
Eq.~(\ref{eqn:substitution}), which is considered as a necessary
requirement from the basic principle of physical invariance. As both
of the two models are built within the framework of effective field
theory in particle physics, they can be considered as two special
cases of a general standard model extension within the effective
field theory. It is therefore necessary to study the relationship
between the two models.

The Lagrangians of the pure photon sector in the SMS and the minimal SME
are
\begin{equation}\label{Lsms}
\mathcal{L}_{\mathrm{SMS}}=-\frac{1}{4}F^{\alpha\beta }F_{\alpha\beta}+ \mathcal{L}_{\mathrm{GV}},
\end{equation}
where
\begin{eqnarray}
\mathcal{L}_{\mathrm{GV}}&=&-\frac{1}{2}\Delta^{\alpha\beta}\Delta^{\mu\nu}(g_{\alpha\mu}\partial_{\beta}
A^{\rho}\partial_{\nu}A_{\rho}-\partial_{\beta}A_{\mu}
\partial_{\nu}A_{\alpha}) \nonumber \\
&&-F_{\mu\nu}\Delta^{\mu\alpha}\partial_{\alpha}A^{\nu},\label{lsms_gv}
\end{eqnarray}
and
\begin{equation}\label{Lsme}
\mathcal{L}_{\mathrm{SME}}=-\frac{1}{4}F^{\alpha\beta }F_{\alpha\beta}+\mathcal{L}_{\mathrm{photon}}^{\mathrm{CPT-even}}
+\mathcal{L}_{\mathrm{photon}}^{\mathrm{CPT-odd}},
\end{equation}
where
\begin{equation}\label{Lsme_e1}
\mathcal{L}_{\mathrm{photon}}^{\mathrm{CPT-even}}=-\frac{1}{4}(k_F)_{\alpha\beta\mu\nu}F^{\alpha\beta}F^{\mu\nu},
\end{equation}
\begin{equation}\label{Lsme_e2}
\mathcal{L}_{\mathrm{photon}}^{\mathrm{CPT-odd}}=\frac{1}{2}(k_{AF})^{\alpha}\epsilon_{\alpha\beta\mu\nu}A^{\beta}F^{\mu\nu}.
\end{equation}
We denote the matrix $\Delta^{\alpha\beta}$ above as Lorentz
invariance violation matrix, whose dimension is massless. When we
ignore the field redefinition, there are 16 independent
dimensionless degrees of freedom in $\Delta^{\alpha\beta}$
generally~\cite{SMS3}.  As coupling constants, the vacuum
expectation value of $\Delta^{\alpha\beta}$ is CPT-even, and the
vacuum expectation value of its derivative
$\partial_{\mu}\Delta^{\alpha\beta}$ is CPT-odd. Coefficients
$(k_F)_{\alpha\beta\mu\nu}$ and $(k_{AF})^{\alpha}$ are CPT-even and
CPT-odd respectively. The CPT-even terms in SME might be understood
as originated from some general relativity consideration as proposed
in Ref.~\cite{LV-GR1}. $(k_F)_{\alpha\beta\mu\nu}$ is antisymmetric
for the first pair indices $\alpha$ and $\beta$, antisymmetric for
the second pair $\mu$ and $\nu$, and symmetric for the interchange
of the two pairs of indices. Hence there are 21 degrees of freedom
in $(k_F)_{\alpha\beta\mu\nu}$. With the redefinition of the gauge
field, there are 2 degrees of freedom to be reduced in
$(k_F)_{\alpha\beta\mu\nu}$. So there are 19 independent degrees of
freedom under the redefinition of the fields and 21 degrees of
freedom in general without considering this redefinition. Another 4
degrees of freedom are in $(k_{AF})^{\alpha}$. After all
consideration, there are $19+4=23$ independent degrees of freedom
for $(k_{AF})^{\alpha}$ and $(k_F)_{\alpha\beta\mu\nu}$ to consider
the field redefinition, and $21+4=25$ ones without considering this
redefinition in general in the pure photon sector of the minimal
SME~\cite{KR08}.  Given the situation that the Lorentz violation
matrix $\Delta^{\alpha\beta}$ here is coupled with other types of
fermions and bosons, there are no universal redefinitions for all
the fields of different particles yet. So we discuss mainly the
general form of $\Delta^{\alpha\beta}$ with all the 16 degrees of
freedom, fit data from various experiments, and obtain the
magnitudes or constraints by the experiments, avoiding any a priori
assumption on $\Delta^{\alpha\beta}$.

We can make a direct correspondence between the two Lagrangians in
Eqs.~(\ref{Lsms}) and (\ref{Lsme}) (cf. the table in
Ref.~\cite{Ma10}), when considering the vacuum expectation values of
both $\Delta^{\alpha\beta}$ (CPT-even) and its derivative
$\partial_{\mu}\Delta^{\alpha\beta}$ (CPT-odd) as coupling constants
and Lorentz violation parameters. In the case that just the Lorentz
violation matrix $\Delta^{\alpha\beta}$ is adopted as the violation
parameters, there are terms left in Eq.~(\ref{Lsms}) which can not
be covered by the Lagrangian in Eqs.~(\ref{Lsme_e1})
and~(\ref{Lsme_e2}). Comparing directly Eq.~(\ref{lsms_gv}) with
Eqs.~(\ref{Lsme_e1}) and~(\ref{Lsme_e2}), we can not find a direct
term-to-term equivalence between the Lagrangians of the free photon
sector in the SMS and the minimal SME.

Here, we treat $\Delta^{\alpha\beta}$ as its vacuum expectation
value, i.e. as coupling constants in the field theory framework.
Then any derivatives $\partial_\mu\Delta^{\alpha\beta}$ vanish in
the following partial integrations during the derivations.

In the standard model supplement, the motion equation for free photons is
\begin{displaymath}
\Pi_{\mathrm{SMS}}^{\gamma\rho}A_{\rho}=0,
\end{displaymath}
and the Lagrangian in Eq.~(\ref{Lsms}) reads also
\begin{equation}\label{Lsms_qaudratic}
\mathcal{L}_{\mathrm{SMS}}=-\frac{1}{2}A_{\gamma}\Pi_{\mathrm{SMS}}^{\gamma\rho}A_{\rho},
\end{equation}
where
\begin{eqnarray}
\Pi_{\mathrm{SMS}}^{\gamma\rho}
&=&-g^{\gamma\rho}\partial^2+\partial^{\gamma}\partial^{\rho}\nonumber\\
&&
+\Delta^{\gamma\alpha}\partial^{\rho}\partial_{\alpha}+\Delta^{\rho\alpha}\partial^{\gamma}\partial_{\alpha}
+\Delta^{\gamma\beta}\Delta^{\rho\nu}\partial_{\beta}\partial_{\nu}\nonumber\\
&&
-g^{\gamma\rho}(2\Delta^{\mu\alpha}\partial_{\mu}\partial_{\alpha}
+g_{\alpha\mu}\Delta^{\alpha\beta}\Delta^{\mu\nu}\partial_{\beta}\partial_{\nu}).
\end{eqnarray}
From Eq.~(\ref{Lsms}) to Eq.~(\ref{Lsms_qaudratic}), partial integrations are used.
We use the Fourier transformation $A_{\rho}(x)=A_{\rho}(p)\exp(-ip\cdot x)$ to get
\begin{eqnarray}
\Pi_{\mathrm{SMS}}^{\gamma\rho}(p)
&=&g^{\gamma\rho}p^2-p^{\gamma}p^{\rho}\nonumber\\
&&
-\Delta^{\gamma\alpha}p^{\rho}p_{\alpha}-\Delta^{\rho\alpha}p^{\gamma}p_{\alpha}
-\Delta^{\gamma\beta}\Delta^{\rho\nu}p_{\beta}p_{\nu}\nonumber\\
&&
+g^{\gamma\rho}(2\Delta^{\mu\alpha}p_{\mu}p_{\alpha}
+g_{\alpha\mu}\Delta^{\alpha\beta}\Delta^{\mu\nu}p_{\beta}p_{\nu})\nonumber.
\end{eqnarray}
For the free photon in the minimal SME, the Lagrangian is similar
\begin{equation}
\mathcal{L}_{\mathrm{SME}}=-\frac{1}{2}A_{\gamma}\Pi_{\mathrm{SME}}^{\gamma\rho}A_{\rho},
\end{equation}
where
\begin{eqnarray}
\Pi_{\mathrm{SME}}^{\gamma\rho}
&=&-g^{\gamma\rho}\partial^2+\partial^{\gamma}\partial^{\rho}
+2(k_F)^{\gamma\alpha\beta\rho}\partial_{\alpha}\partial_{\beta}\nonumber\\
&&+2(k_{AF})_{\alpha}\epsilon^{\gamma\alpha\beta\rho}\partial_{\beta},
\end{eqnarray}
and the representation in momentum space is
\begin{eqnarray}
\Pi_{\mathrm{SME}}^{\gamma\rho}(p)
&=&g^{\gamma\rho}p^2-p^{\gamma}p^{\rho}
-2(k_F)^{\gamma\alpha\beta\rho}p_{\alpha}p_{\beta}\nonumber\\
&&-2i(k_{AF})_{\alpha}\epsilon^{\gamma\alpha\beta\rho}p_{\beta}.\nonumber
\end{eqnarray}
We see that $\Pi_{\mathrm{SMS}}^{\gamma\rho}(p)$ and $\Pi_{\mathrm{SME}}^{\gamma\rho}(p)$
are the inverse of the photon propagator in the momentum space. The propagator determines
the propagating properties of photons.

When the two Lagrangians Eq.~(\ref{Lsms}) and Eq.~(\ref{Lsme}) are
equivalent to each other for free photons, i.e. we consider the
common part (intersection) between the two models, some
enlightenments are expected to come. We can get
$\Pi_{\mathrm{SMS}}^{\gamma\rho}=\Pi_{\mathrm{SME}}^{\gamma\rho}$.
Then the matrix equation is satisfied
\begin{eqnarray}
&&g^{\gamma\rho}(2\Delta^{\mu\alpha}p_{\mu}p_{\alpha}
+g_{\alpha\mu}\Delta^{\alpha\beta}\Delta^{\mu\nu}p_{\beta}p_{\nu})\nonumber\\
&&
-\Delta^{\gamma\alpha}p^{\rho}p_{\alpha}-\Delta^{\rho\alpha}p^{\gamma}p_{\alpha}
-\Delta^{\gamma\beta}\Delta^{\rho\nu}p_{\beta}p_{\nu}\nonumber\\
&=&
-2(k_F)^{\gamma\alpha\beta\rho}p_{\alpha}p_{\beta}
-2i(k_{AF})_{\alpha}\epsilon^{\gamma\alpha\beta\rho}p_{\beta}\nonumber.
\end{eqnarray}
Making derivative with respect to momentum $p^{\alpha}$ for two
times, we obtain
\begin{eqnarray}
&&2g_{\gamma\rho}(\Delta_{\alpha\beta}+\Delta_{\beta\alpha}
+g^{\mu\nu}\Delta_{\mu\alpha}\Delta_{\nu\beta})\nonumber\\
&&-\Delta_{\gamma\alpha}g_{\rho\beta}-\Delta_{\gamma\beta}g_{\rho\alpha}
-\Delta_{\rho\alpha}g_{\gamma\beta}-\Delta_{\rho\beta}g_{\gamma\alpha}\nonumber\\
&&
-\Delta_{\gamma\alpha}\Delta_{\rho\beta}-\Delta_{\gamma\beta}\Delta_{\rho\alpha}\nonumber\\
&=&
-4(k_F)_{\gamma(\alpha\beta)\rho}
-4i(k_{AF})^{\mu}\epsilon_{\mu\gamma\rho(\alpha}l_{\beta)}^s.
\end{eqnarray}
We accept conventions of general relativity for the notation
of indices here and in the following derivations.
The coefficient $l_{\beta}^s$ is introduced here, and
its dimension is $[\mathrm{length}]$ or $[\mathrm{mass}]^{-1}$.
$l_{\beta}^s$ represents the characteristic length of the physical process.
Based on the symmetric/antisymmetric properties of indices
$\gamma$ and $\rho$, we get two matrix equations further
\begin{equation}\label{cs1}
-i(k_{AF})^{\mu}\epsilon_{\mu\gamma\rho(\alpha}l_{\beta)}^s=(k_F)_{[\gamma|(\alpha\beta)|\rho]}=0,
\end{equation}
and
\begin{eqnarray}\label{cs2}
&&2g_{\gamma\rho}(\Delta_{\alpha\beta}+\Delta_{\beta\alpha}
+g^{\mu\nu}\Delta_{\mu\alpha}\Delta_{\nu\beta})\nonumber\\
&&-\Delta_{\gamma\alpha}g_{\rho\beta}-\Delta_{\gamma\beta}g_{\rho\alpha}
-\Delta_{\rho\alpha}g_{\gamma\beta}-\Delta_{\rho\beta}g_{\gamma\alpha}\nonumber\\
&&
-\Delta_{\gamma\alpha}\Delta_{\rho\beta}-\Delta_{\gamma\beta}\Delta_{\rho\alpha}\nonumber\\
&=&
-4(k_F)_{(\gamma|(\alpha\beta)|\rho)}=-4(k_F)_{\gamma(\alpha\beta)\rho}.
\end{eqnarray}
The general formula Eq.~(\ref{cs2}) here demonstrates the relations
of the Lorentz violation matrix $\Delta_{\alpha\beta}$ and the
coefficient $(k_F)_{\gamma\alpha\beta\rho}$, for the intersection of
the SMS and the minimal SME.

When we take $k_{AF}$ and $k_F$ of Eqs.~(\ref{cs1}) and (\ref{cs2})
into the Lagrangians of the minimal SME of Eqs.~(\ref{Lsme_e1})
and~(\ref{Lsme_e2}), we find that the Lagrangian of the minimal SME
of Eq.~(\ref{Lsme}) can be converted to that of the SMS of
Eq.~(\ref{Lsms}). This tells us that the free photon sector of the SMS can be considered as a
subset of the minimal SME, provided that the coefficients
$(k_{AF})^\mu$ and $(k_F)_{\gamma\alpha\beta\rho}$
are constrained by Eqs.~(\ref{cs1}) and (\ref{cs2}).

The identity Eq.~(\ref{cs1}) tells us the relations between
$(k_F)_{\gamma\alpha\beta\rho}$ and $(k_{AF})^{\mu}$. The
constraints  mean that the violation coefficient $(k_{AF})^{\mu}$
(CPT-odd) vanishes in the photon sector of the minimal SME, i.e.
\begin{displaymath}
(k_{AF})^{\mu}=0.
\end{displaymath}
There is a maximal sensitivity $10^{-42}\sim10^{-43}$~GeV for the
coefficients $k^{(3)}_{(V)00}$, $k^{(3)}_{(V)10}$, $\mathrm{Re}
k^{(3)}_{(V)11}$ and $\mathrm{Im} k^{(3)}_{(V)11}$ in
Tab.~\ref{t_comparison}. These four parameters are defined in terms
of coefficient $(k_{AF})^{\alpha}$ in Tab.~\ref{t_relation}. In
Eq.~(\ref{cs2}), tensor $(k_F)_{\gamma\alpha\beta\rho}$ is
antisymmetric for indices $\gamma,\alpha$, and antisymmetric for
indices $\beta,\rho$. At the same time,
$(k_F)_{\gamma\alpha\beta\rho}=(k_F)_{\beta\rho\gamma\alpha}$. So
there are 21 degrees of freedom in $(k_F)_{\gamma\alpha\beta\rho}$
generally, and $21-6=15$ independent elements in tensor
$(k_F)_{\gamma(\alpha\beta)\rho}$. We have already known that there
are 16 degrees of freedom for $\Delta^{\alpha\beta}$, without the
field redefinitions being considered. Therefore, any one of
$\Delta^{\alpha\beta}$ and $(k_F)_{\gamma\alpha\beta\rho}$ can not
completely determines the other one. In Eq.~(\ref{cs2}), there are
15 degrees of freedom of the Lorentz violation matrix
$\Delta_{\alpha\beta}$ related with 15 ones of the tensor
$(k_F)_{\gamma\alpha\beta\rho}$. A definite $\Delta^{\alpha\beta}$
can determine at most 15 degrees of freedom  of
$(k_F)_{\gamma\alpha\beta\rho}$, and vice versa.

We consider an example that $\Delta_{\alpha\beta}$ is a symmetric matrix.
Then we can get the explicit form for it. Multiplying $g^{\gamma\rho}$ on
both sides of Eq.~(\ref{cs2}), we obtain
\begin{displaymath}
3g^{\mu\nu}\Delta_{\mu\alpha}\Delta_{\nu\beta}+6\Delta_{(\alpha\beta)}=-2(k_F)_{\alpha\beta},
\end{displaymath}
where we define $(k_F)_{\alpha\beta}=g^{\gamma\rho}(k_F)_{\gamma\alpha\beta\rho}$.
The tensor $(k_F)_{\gamma\alpha\beta\rho}$ has most of the properties of the
Riemann curvature tensor. So $(k_F)_{\alpha\beta}$ is a ``Ricci tensor" and satisfies that
$(k_F)_{\alpha\beta}=(k_F)_{\beta\alpha}$. As we know that $(k_F)_{\alpha\beta}\ll 1$ and
$\Delta_{\alpha\beta}\ll 1$, the solution of $\Delta_{\alpha\beta}$ in terms of
$(k_F)_{\alpha\beta}$, to the second order, is
\begin{equation}
\Delta_{\alpha\beta}=-\frac{1}{3}(k_F)_{\alpha\beta}-\frac{1}{18}g^{\mu\nu}(k_F)_{\mu\alpha}(k_F)_{\nu\beta}.
\end{equation}
With the assumption of $\Delta_{\alpha\beta}$ being a symmetric
matrix, there are 10 independent elements in it. Then the Lorentz
invariance violation matrix for the free photon can be obtained from
the tensor $(k_F)_{\gamma\alpha\beta\rho}$ completely, and can be
considered as somewhat a kind of Ricci tensor. In the following
part, the general case of $\Delta_{\alpha\beta}$ with all the 16
degrees of freedom is considered.

\section{Lorentz Violation Matrix of Photons}\label{sec_photon_lvm}
The Lorenz violation parameters of the minimal SME have been
constrained from various recent experiments. The data can also
provide bounds on the magnitudes of the elements of the Lorentz
violation matrix of photons in the SMS, through Eqs.~(\ref{cs1})
and~(\ref{cs2}) above. The Lorentz violation parameters of the SME
commonly used in experiments are four matrices
$(\tilde{\kappa}_{e+})^{jk}$, $(\tilde{\kappa}_{e-})^{jk}$,
$(\tilde{\kappa}_{o+})^{jk}$ and $(\tilde{\kappa}_{o-})^{jk}$ in
Ref.~\cite{KR08}:
\begin{displaymath}
(\tilde{\kappa}_{e+})^{jk}=-(k_F)^{0j0k}+\frac{1}{4}\epsilon^{jpq}\epsilon^{krs}(k_F)^{pqrs},
\end{displaymath}
\begin{displaymath}
(\tilde{\kappa}_{e-})^{jk}=-(k_F)^{0j0k}-\frac{1}{4}\epsilon^{jpq}\epsilon^{krs}(k_F)^{pqrs}+\frac{2}{3}(k_F)^{0l0l}\delta^{jk},
\end{displaymath}
\begin{displaymath}
(\tilde{\kappa}_{o+})^{jk}=-\frac{1}{2}\epsilon^{jpq}(k_F)^{0kpq}+\frac{1}{2}\epsilon^{kpq}(k_F)^{0jpq},
\end{displaymath}
and
\begin{displaymath}
(\tilde{\kappa}_{o-})^{jk}=\frac{1}{2}\epsilon^{jpq}(k_F)^{0kpq}+\frac{1}{2}\epsilon^{kpq}(k_F)^{0jpq}.
\end{displaymath}
The same indices mean summation. More parameters related with
the Lorentz violation matrix are listed in Tab.~\ref{t_relation}.

In terms of $(k_F)^{\alpha\beta\mu\nu}$, the four matrices can be
rewritten as follows,
\begin{widetext}
\begin{displaymath}
(\tilde{\kappa}_{e+})^{jk}=
\left(
  \begin{array}{ccc}
    -(k_F)^{0101}+(k_F)^{2323} & -(k_F)^{0102}+(k_F)^{2331} & -(k_F)^{0103}+(k_F)^{2312} \\
    " & -(k_F)^{0202}+(k_F)^{3131} & -(k_F)^{0203}+(k_F)^{3112} \\
    " & " & -(k_F)^{0303}+(k_F)^{1212} \\
  \end{array}
\right),
\end{displaymath}
which is symmetric for the indices $j$ and $k$;
\begin{displaymath}
(\tilde{\kappa}_{e-})^{jk}=
\left(
  \begin{array}{ccc}
    -(k_F)^{0101}-(k_F)^{2323}+\alpha & -(k_F)^{0102}-(k_F)^{2331} & -(k_F)^{0103}-(k_F)^{2312} \\
    " & -(k_F)^{0202}-(k_F)^{3131}+\alpha & -(k_F)^{0203}-(k_F)^{3112} \\
    " & " & -(k_F)^{0303}-(k_F)^{1212}+\alpha \\
  \end{array}
\right),
\end{displaymath}
which is symmetric for the indices $j$ and $k$, and
$\alpha\equiv\frac{2}{3}(k_F)^{0l0l}$;
\begin{displaymath}
(\tilde{\kappa}_{o+})^{jk}=
\left(
  \begin{array}{ccc}
    0 & (k_F)^{0131}-(k_F)^{0223} & (k_F)^{0112}-(k_F)^{0323} \\
    " & 0 & (k_F)^{0212}-(k_F)^{0331} \\
    " & " & 0 \\
  \end{array}
\right),
\end{displaymath}
which is antisymmetric for the indices $j$ and $k$; and
\begin{displaymath}
(\tilde{\kappa}_{o-})^{jk}=
\left(
  \begin{array}{ccc}
    2(k_F)^{0123} & (k_F)^{0131}+(k_F)^{0223} & (k_F)^{0112}+(k_F)^{0323} \\
    " & 2(k_F)^{0231} & (k_F)^{0212}+(k_F)^{0331} \\
    " & " & 2(k_F)^{0312} \\
  \end{array}
\right),
\end{displaymath}
\end{widetext}
which is symmetric for the indices $j$ and $k$.
The symbol " above means that the matrix element is not written
explicitly for brevity and it can be obtained from the property of
symmetry/antisymmetry of the corresponding matrix.

With Eq.~(\ref{cs2}), we can replace the tensor
$(k_F)^{\alpha\beta\mu\nu}$ with the Lorentz violation matrix
$\Delta^{\alpha\beta}$. So the four matrices above are rewritten
like:
\begin{widetext}
\begin{equation}\label{ke1_d}
(\tilde{\kappa}_{e+})^{jk}=
\left( \begin {array}{ccc}
\frac{1}{2}\Delta & \frac{1}{2}(\Delta^{10}\Delta^{20}+\Delta^{23}\Delta^{13})&\frac{1}{2}(\Delta^{10}\Delta^{30}
+\Delta^{32}\Delta^{12})\\
" & \frac{1}{2}\Delta & \frac{1}{2}(\Delta^{20}\Delta^{30}+\Delta^{31}\Delta^{21})\\
" & " & \frac{1}{2}\Delta
\end {array} \right),
\end{equation}
which is symmetric for the indices $j$ and $k$, with
$\Delta=\Delta^{\alpha\beta}g_{\alpha\beta}$;
\begin{equation}\label{ke0_d}
(\tilde{\kappa}_{e-})^{jk}=
\left( \begin {array}{ccc} \frac{1}{2}\Delta^{00}+\frac{1}{6}\Delta^{11}-\frac{5}{6}\Delta^{22}-\frac{5}{6}\Delta^{33}&\frac{1}{2}(\Delta^{10}\Delta^{20}
-\Delta^{23}\Delta^{13})& \frac{1}{2}(\Delta^{10}\Delta^{30}-\Delta^{32}\Delta^{12})\\
" &\frac{1}{2}\Delta^{00}+\frac{1}{6}\Delta^{22}-\frac{5}{6}\Delta^{11}-\frac{5}{6}\Delta^{33}&\frac{1}{2}
(\Delta^{20}\Delta^{30}-\Delta^{31}\Delta^{21})\\
" & " &\frac{1}{2}\Delta^{00}+\frac{1}{6}\Delta^{33}-\frac{5}{6}\Delta^{11}-\frac{5}{6}\Delta^{22}
\end {array} \right),
\end{equation}
which is symmetric for the indices $j$ and $k$, with
$\alpha\equiv\frac{2}{3}(k_F)^{0l0l}$;
\begin{equation}\label{ko1_d}
(\tilde{\kappa}_{o+})^{jk}=
\left( \begin {array}{ccc}
0 &-\frac{1}{2}(\Delta^{02}\Delta^{32}+\Delta^{01}\Delta^{31})&\frac{1}{2}(\Delta^{03}\Delta^{23}+\Delta^{01}\Delta^{21})\\
" & 0 &-\frac{1}{2}(\Delta^{03}\Delta^{13}+\Delta^{02}\Delta^{12})\\
" & " & 0
\end {array} \right),
\end{equation}
which is antisymmetric for the indices $j$ and $k$; and
\begin{equation}\label{ko0_d}
(\tilde{\kappa}_{o-})^{jk}=
\left( \begin {array}{ccc}
? & \frac{1}{2}(\Delta^{02}\Delta^{32}-\Delta^{01}\Delta^{31}) &\frac{1}{2}(\Delta^{01}\Delta^{21}-\Delta^{03}\Delta^{23})\\
" & ? &\frac{1}{2}(\Delta^{03}\Delta^{13}-\Delta^{02}\Delta^{12})\\
" & " & ?
\end {array} \right),
\end{equation}
\end{widetext}
which is symmetric for the indices $j$ and $k$. The metric tensor
$g_{\alpha\beta}=\mathrm{diag}(-1,1,1,1)$ is used here. The symbol
`?' in matrix $(\tilde{\kappa}_{o-})^{jk}$ denotes that the involved
elements $(k_F)^{0123}$, $(k_F)^{0231}$ and $(k_F)^{0312}$ can not
be written in terms of the Lorentz violation matrix
$\Delta^{\alpha\beta}$ through Eq.~(\ref{cs2}). We have known from
the preceding section that only 15 degrees of freedom in matrix
$\Delta^{\alpha\beta}$ are inter-related with the same number of
independent degrees of freedom in tensor
$(k_F)_{\gamma\alpha\beta\rho}$, and vice versa. Besides the
representation of $(\tilde{\kappa}_{e+})^{jk}$,
$(\tilde{\kappa}_{e-})^{jk}$, $(\tilde{\kappa}_{o+})^{jk}$ and
$(\tilde{\kappa}_{o-})^{jk}$ in terms of $\Delta^{\alpha\beta}$, the
relations between other Lorentz violation parameters commonly used
in the minimal SME and the Lorentz violation matrix here are
summarized in Tab.~\ref{t_relation}. There are 15 independent
expressions in terms of $\Delta^{\alpha\beta}$ appearing in the four
matrices $(\tilde{\kappa}_{e+})^{jk}$,
$(\tilde{\kappa}_{e-})^{jk}$, $(\tilde{\kappa}_{o+})^{jk}$ and
$(\tilde{\kappa}_{o-})^{jk}$.  These 15 independent expressions help
to determine 15 degrees of freedom of the Lorentz violation matrix
$\Delta^{\alpha\beta}$.

With the recent maximal sensitivities attained from current experiments for Lorentz violation parameters
$(\tilde{\kappa}_{e+})^{jk}$, $(\tilde{\kappa}_{e-})^{jk}$, $(\tilde{\kappa}_{o+})^{jk}$ and
$(\tilde{\kappa}_{o-})^{jk}$ of the free photon sector in the minimal SME (see Tab.~\ref{t_comparison}),
we get the maximal sensitivities or the conservative bounds from experiments for
Lorentz invariance matrix $\Delta^{\alpha\beta}_{\mathrm{photon}}$
\begin{equation}\label{lvmp}
\left( \begin{array}{cccc}
3\Delta^{33}+10^{-17}&10^{-5}&10^{-5}&10^{-6}\\
10^{-9}&\Delta^{33}+10^{-17}&10^{-9}&10^{-9}\\
10^{-9}&10^{-9}&\Delta^{33}+10^{-17}&10^{-9}\\
10^{-9}&10^{-8}&10^{-8}&\Delta^{33}
\end{array} \right),
\end{equation}
in the Sun-centered inertial reference frame~\cite{KR08,BKLR03}.
The publication~\cite{KR08} claimed a $2\sigma$ limit on Lorentz violation coefficients
$(\tilde{\kappa}_{e+})^{jk}$, $(\tilde{\kappa}_{e-})^{jk}$, $(\tilde{\kappa}_{o+})^{jk}$,
$(\tilde{\kappa}_{o-})^{jk}$ and etc. in Tab.~\ref{t_comparison}.
The 15 independent degrees of freedom of $\Delta^{\alpha\beta}$ are determined,
and there is still one freedom $\Delta^{33}$ remaining unclear. The maximal sensitivity
for the elements of $\Delta^{\alpha\beta}$ is listed in Tab.~\ref{t_delta}.

Eq.~(\ref{lvmp}) demonstrates that the Lorentz violation matrix
$\Delta^{\alpha\beta}$ does not need to be a symmetric matrix in general.
The non-symmetric structure of the photon Lorentz violation matrix
suggests preferred directions and potential anisotropy of
space-time~\cite{Ma10graal,Bocquet10} for propagating of the free
photon, even in the case of no gravitation. More experiments will
give more details for $\Delta^{\alpha\beta}$.

There are different representations for the Lorentz violation matrix
$\Delta^{\alpha\beta}$ in different coordinate systems. These
representations are related with each other by a coordinate
transformation matrix $T^{\alpha\beta}$ in group SO(1,3). When the
relative velocity between these two coordinate systems is much
smaller than the light speed, the element of $T^{\alpha\beta}$ is
either order $O(1)$ or close to zero. An element of the Lorentz
violation matrix in a coordinate system is the linear combinations
of the elements of $\Delta^{\alpha\beta}$ in the other coordinate
system. Then the magnitudes of the Lorentz violation matrix in these
two coordinates are not different too much from each other. So we
expect that the upper bound on the violation parameters appearing in
Ref.~\cite{Ma10graal} is compatible with the maximal sensitivity
shown in Eq.~(\ref{lvmp}). The limit of order $10^{-14}$ in
Ref.~\cite{Ma10graal} for the photon Lorentz violation matrix is
indeed compatible with the bound $10^{-8}$  here.

From Eq.~(\ref{lvmp}), we find that the trace $\Delta\equiv
\mathrm{tr}(\Delta^{\alpha\beta})=g_{\alpha\beta}\Delta^{\alpha\beta}\simeq10^{-17}$.
A competitive upper bound $1.6\times 10^{-14}$ on the photon Lorentz
matrix $\Delta^{\alpha\beta}$ was obtained in Ref.~\cite{Ma10graal}.
In that article~\cite{Ma10graal}, we made an assumption about the
form of the matrix $\Delta^{\alpha\beta}$ theoretically for the
analysis on the data there. There is no a priori assumptions here about
the general structure of $\Delta^{\alpha\beta}$. We see that the
maximal attained sensitivity $10^{-17}$ for the trace of the Lorentz
violation matrix is stronger than the upper limit $10^{-14}$ gotten
in Ref.~\cite{Ma10graal}. Compared with the stringent bound on the
trace $\Delta$, the maximal attained sensitivities put looser limits
on the non-diagonal elements of $\Delta^{\alpha\beta}$, shown in
Eq.~(\ref{lvmp}).

Through this work, we have seen that the two theories of the SMS and
the minimal SME can give same results for free photons.
Eq.~(\ref{cs2}) shows the correlations between the Lorentz
invariance violation matrix $\Delta^{\alpha\beta}$ of our model and
the coupling tensor $(k_F)_{\alpha\beta\mu\nu}$ appearing in the
photon sector of the minimal SME. The relations of the violation
parameters $\Delta^{\alpha\beta}$ with the parameters
$(k_F)_{\alpha\beta\mu\nu}$ uncover the detailed structure of the
Lorentz violation matrix of free photons in Eq.~(\ref{lvmp}).
Up to now, there have been no compelling experimental evidences for the
existence of Lorentz violation for photons. All that we have gotten
so far are the theoretical analysis and the maximal sensitivities
attained from the recent experiments.

\section{Conclusion}
Two Lorentz violation models, the minimal standard model extension
(SME) and the standard model supplement (SMS), are compared here for
the photon sector. For all the terms in the Lagrangians of the pure
photon sector, there is no direct one-to-one correspondence between
the two models in general. However, some interesting results can be
obtained by the requirement that the two models are identical with
each other in the intersection. We find that the free photon sector of the SMS can be a
subset of the minimal SME provided with some connections in the SME
parameters. (i) We consider the photon sector of the two models, and
two main equations are obtained between $\Delta^{\alpha\beta}$,
$(k_F)^{\alpha\beta\mu\nu}$ and $(k_{AF})^{\alpha}$, through the
propagator of photons in the momentum space. (ii) These equations
suggest that the CPT-odd coefficients $(k_{AF})^{\alpha}$ are zero.
Such a suggestion is supported by available experimental bounds, for
example, there is the maximal sensitivity $10^{-42}\sim10^{-43}$~GeV
from experiments for the coefficients $k^{(3)}_{(V)00}$,
$k^{(3)}_{(V)10}$, $\mathrm{Re} k^{(3)}_{(V)11}$ and $\mathrm{Im}
k^{(3)}_{(V)11}$. (iii) There are 15 degrees of freedom in the
Lorentz violation matrix $\Delta^{\alpha\beta}$ and the same number
of degrees of freedom in tensor $(k_F)^{\alpha\beta\mu\nu}$ to be
inter-related. We got the conservative bound Eq.~(\ref{lvmp}) on the
detailed structure of the photon Lorentz violation matrix in our
model. The bounds on $\Delta^{\alpha\beta}$ are gotten from the
limits on the Lorentz parameters of the minimal SME. The detailed
structure of the photon Lorentz violation matrix can play an
important role for applications to Lorentz violation experiments.

For $\Delta^{\alpha\beta}$ of free photons, due to the factor that a
universal constant can be absorbed into the gauge field $A^\mu$,
there are 15 independent degrees of freedom in
$\Delta^{\alpha\beta}$ of free photons to describe Lorentz
violation. In the paper, we do not use these 15 independent degrees
of freedom to derive the magnitudes of all the 16 elements of
$\Delta^{\alpha\beta}$, but use the relations of it with the
parameters in the minimal SME to get the constraints on
$\Delta^{\alpha\beta}$ of free photons from the constraints of
various experiments on the minimal SME.


The strong constraints on the matrix elements of
$\Delta^{\alpha\beta}$ mean that Lorentz violation is small for
photons if it exists. The matrix $\Delta^{\alpha\beta}$ is not
symmetric generally. The non-symmetry property of
$\Delta^{\alpha\beta}$ implies that the space-time for free photons
can be not isotropic very well, even if in the case of no gravitation.
To date, there has been theoretical analysis on Lorentz
violation and there is no strong experimental evidence supporting
Lorentz violation for photons. Generally, we should study the
minimal standard model extension and the standard model supplement
separately, and then determine whether these two models are
equivalent to each other by directly confronting  relevant
experiments.

\begin{acknowledgements}
The work was supported by National Natural Science Foundation of
China (Grants No.~11035003 and No.~11120101004) and
by the Research Fund for the Doctoral Program of Higher Education,
China.
\end{acknowledgements}

\appendix
\section{Tables}

\begin{table*}
  \centering
  \caption{Maximal sensitivities ($2\sigma$) for the Lorentz violation matrix of photons.}\label{t_delta}
\begin{tabular}{cc}
  \hline
  Coefficient & Sensitivity  \\
  \hline
  $\Delta^{00}-3\Delta^{33}$ & $10^{-17}$ \\
  $\Delta^{11}-\Delta^{33}$ & $10^{-17}$ \\
  $\Delta^{22}-\Delta^{33}$ & $10^{-17}$ \\\\
  $\Delta^{01}$ & $10^{-5}$ \\
  $\Delta^{02}$ & $10^{-5}$ \\
  $\Delta^{03}$ & $10^{-6}$ \\\\
  $\Delta^{10}$ & $10^{-9}$ \\
  $\Delta^{20}$ & $10^{-9}$ \\
  $\Delta^{30}$ & $10^{-9}$ \\\\
  $\Delta^{12}$ & $10^{-9}$ \\
  $\Delta^{13}$ & $10^{-9}$ \\
  $\Delta^{21}$ & $10^{-9}$ \\
  $\Delta^{23}$ & $10^{-9}$ \\
  $\Delta^{31}$ & $10^{-8}$ \\
  $\Delta^{32}$ & $10^{-8}$ \\
  \hline
\end{tabular}
\end{table*}

\begin{table*}
  \centering
  \caption{Maximal sensitivities ($2\sigma$) for the photon sector (from Ref.~\cite{KR08}). The superscripts $X,Y,Z$ there are
  converted to $1,2,3$ here respectively for consistence with the notation of the Lorentz violation matrix.}\label{t_comparison}
\begin{tabular}{cc}
  \hline
  Coefficient & Sensitivity  \\
  \hline
  $(\tilde{\kappa}_{e+})^{12}$ & $10^{-32}$ \\
  $(\tilde{\kappa}_{e+})^{13}$ & $10^{-32}$ \\
  $(\tilde{\kappa}_{e+})^{23}$ & $10^{-32}$ \\
  $(\tilde{\kappa}_{e+})^{11}-(\tilde{\kappa}_{e+})^{22}$ & $10^{-32}$ \\
  $(\tilde{\kappa}_{e+})^{33}$ & $10^{-32}$ \\\\
  $(\tilde{\kappa}_{o-})^{12}$ & $10^{-32}$ \\
  $(\tilde{\kappa}_{o-})^{13}$ & $10^{-32}$ \\
  $(\tilde{\kappa}_{o-})^{23}$ & $10^{-32}$ \\
  $(\tilde{\kappa}_{o-})^{11}-(\tilde{\kappa}_{o-})^{22}$& $10^{-32}$ \\
  $(\tilde{\kappa}_{o-})^{33}$ & $10^{-32}$ \\\\
  $(\tilde{\kappa}_{e-})^{12}$ & $10^{-17}$ \\
  $(\tilde{\kappa}_{e-})^{13}$ & $10^{-17}$ \\
  $(\tilde{\kappa}_{e-})^{23}$ & $10^{-17}$ \\
  $(\tilde{\kappa}_{e-})^{11}-(\tilde{\kappa}_{e-})^{22}$ & $10^{-17}$ \\
  $(\tilde{\kappa}_{e-})^{33}$ & $10^{-16}$ \\\\
  $(\tilde{\kappa}_{o+})^{12}$ & $10^{-13}$ \\
  $(\tilde{\kappa}_{o+})^{13}$ & $10^{-14}$ \\
  $(\tilde{\kappa}_{o+})^{23}$ & $10^{-14}$ \\\\
  $\tilde{\kappa}_{\mathrm{tr}}$ & $10^{-14}$ \\\\
  $k^{(3)}_{(V)00}$ & $10^{-43}$~GeV \\
  $k^{(3)}_{(V)10}$ & $10^{-42}$~GeV \\
  $\mathrm{Re} k^{(3)}_{(V)11}$ & $10^{-42}$~GeV \\
  $\mathrm{Im} k^{(3)}_{(V)11}$ & $10^{-42}$~GeV \\
  \hline
\end{tabular}
\end{table*}

\begin{table*}
  \centering
  \caption{Definitions for the photon sector in the minimal SME, together with relations with the Lorentz violation matrix of the SMS.
  (The two left columns are from Ref.~\cite{KR08}.)}\label{t_relation}
\begin{tabular}{ccc}
  \hline
  Symbol & Definition & Relation \\ \hline
  $(\tilde{\kappa}_{e+})^{jk}$ & $-(k_F)^{0j0k}+\frac{1}{4}\epsilon^{jpq}\epsilon^{krs}(k_F)^{pqrs}$ & Eq. (\ref{ke1_d}) \\
  $(\tilde{\kappa}_{e-})^{jk}$ & $-(k_F)^{0j0k}-\frac{1}{4}\epsilon^{jpq}\epsilon^{krs}(k_F)^{pqrs}+\frac{2}{3}(k_F)^{0l0l}\delta^{jk}$
  & Eq. (\ref{ke0_d}) \\
  $(\tilde{\kappa}_{o+})^{jk}$ & $-\frac{1}{2}\epsilon^{jpq}(k_F)^{0kpq}+\frac{1}{2}\epsilon^{kpq}(k_F)^{0jpq}$  & Eq. (\ref{ko1_d}) \\
  $(\tilde{\kappa}_{o-})^{jk}$ & $\frac{1}{2}\epsilon^{jpq}(k_F)^{0kpq}+\frac{1}{2}\epsilon^{kpq}(k_F)^{0jpq}$ & Eq. (\ref{ko0_d}) \\
  $\tilde{\kappa}_{\mathrm{tr}}$ & $-\frac{2}{3}(k_F)^{0l0l}$ & $-\frac{2}{3}\Delta^{11}+\frac{1}{3}\Delta$ \\\\
  $k^1$ & $(k_F)^{0213}$ & ? \\
  $k^2$ & $(k_F)^{0123}$ & ? \\
  $k^3$ & $(k_F)^{0202}-(k_F)^{1313}$ & $-\frac{1}{2}\Delta$ \\
  $k^4$ & $(k_F)^{0303}-(k_F)^{1212}$ & $-\frac{1}{2}\Delta$ \\
  $k^5$ & $(k_F)^{0102}+(k_F)^{1323}$ & $-\frac{1}{2}(\Delta^{10}\Delta^{20}+\Delta^{23}\Delta^{13})$ \\
  $k^6$ & $(k_F)^{0103}-(k_F)^{1223}$ & $-\frac{1}{2}(\Delta^{10}\Delta^{30}+\Delta^{32}\Delta^{12})$ \\
  $k^7$ & $(k_F)^{0203}+(k_F)^{1213}$ & $-\frac{1}{2}(\Delta^{20}\Delta^{30}+\Delta^{31}\Delta^{21})$ \\
  $k^8$ & $(k_F)^{0112}+(k_F)^{0323}$ & $-\frac{1}{2}(\Delta^{03}\Delta^{23}-\Delta^{01}\Delta^{21})$ \\
  $k^9$ & $(k_F)^{0113}-(k_F)^{0223}$ & $\frac{1}{2}(\Delta^{01}\Delta^{31}-\Delta^{02}\Delta^{32})$ \\
  $k^{10}$ & $(k_F)^{0212}-(k_F)^{0313}$ & $\frac{1}{2}(\Delta^{03}\Delta^{13}-\Delta^{02}\Delta^{12})$ \\\\
  $k^{(3)}_{(V)00}$ & $-\sqrt{4\pi}(k_{AF})^0$ & 0\\
  $k^{(3)}_{(V)10}$ & $-\sqrt{4\pi/3}(k_{AF})^3$  & 0\\
  $\mathrm{Re} k^{(3)}_{(V)11}$ & $\sqrt{2\pi/3}(k_{AF})^1$ & 0\\
  $\mathrm{Im} k^{(3)}_{(V)11}$ & $-\sqrt{2\pi/3}(k_{AF})^2$ & 0\\
  \hline
\end{tabular}
\end{table*}


\begin{thebibliography}{00}


\bibitem{AmelinoCamelia:2008qg}
For a recent review on quantum gravity theories and the relevant Lorentz violation studies, see, e.g.,
  G.~Amelino-Camelia,
  ``Quantum Gravity Phenomenology,''
  Living Rev.\ Rel.\  {16}, 5 (2013),
    and references therein.

\bibitem{ShaoMa10}
For a brief review on Lorentz violation effects through very high
energy photons of astrophysical sources, see. e.g.,
  L.~Shao, B.-Q.~Ma,
  ``Lorentz violation effects on astrophysical propagation of very high energy
  photons,''
Mod.\ Phys.\ Lett.\  A {\bf 25}, 3251 (2010) [arXiv:1007.2269], and
references therein.

\bibitem{Dirac}
P.A.M. Dirac,
Is there an {\AE}ther,
Nature {\bf 168}, 906 (1951).

\bibitem{Bjorken}
J.D. Bjorken,
A dynamical origin for the electromagnetic field,
Ann.\ Phys. {\bf 24}, 174 (1963).

\bibitem{p99}
M. Planck, $\ddot{\rm U}$ber irreversible
  Strahlungsvorg$\ddot{\rm a}$nge,
  Sitzungsberichte der K$\ddot{\rm o}$niglich Preu$\ss$ischen Akademie der Wissenschaften zu Berlin {\bf 5}, 440 (1899).

\bibitem{Snyder}
H.S. Snyder, Quantized space-time, Phys. Rev. 71 (1947) 38.

\bibitem{Snyder1}
H.S. Snyder, The electromagnetic field in quantized spacetime, Phys. Rev. 72 (1947) 68.

\bibitem{xu-l}
  Y.~Xu and B.-Q.~Ma,
 ``Universal entropy bound and discrete space-time,''
  Mod.\ Phys.\ Lett.\  A {\bf 26}, 2101 (2011)
  [arXiv:1106.1778 [hep-th]].

\bibitem{lv5}
  L.~Shao, B.-Q.~Ma,
  ``Note on a new fundamental length scale $l$ instead of the Newtonian constant $G$,''
  Sci. China Phys. Mech. Astro. {\bf 54}, 1771 (2011)
  [arXiv:1006.3031 [hep-th]].




\bibitem{Amelino-Camelia2002}
  G.~Amelino-Camelia,
  ``Relativity in space-times with short-distance structure governed by an
  observer-independent (Planckian) length scale,''
  Int.\ J.\ Mod.\ Phys.\  D {\bf 11}, 35 (2002)
  [arXiv:gr-qc/0012051].

\bibitem{Magueijo:2001cr}
  J.~Magueijo and L.~Smolin,
  ``Lorentz invariance with an invariant energy scale,''
  Phys.\ Rev.\ Lett.\  {\bf 88}, 190403 (2002)
  [arXiv:hep-th/0112090].

\bibitem{Zhang2011}
  X.~Zhang, L.~Shao, B.-Q.~Ma,
  ``Photon Gas Thermodynamics in Doubly Special Relativity,''
  Astropart.\ Phys.\  {\bf 34}, 840 (2011).

\bibitem{LV-GR1}
W.-T. Ni, Yang¡¯s gravitational field equations, Phys.\ Rev.\ Lett.\ {\bf 35}, 319 (1975).
\bibitem{LV-GR10}
  W.-T. Ni,
  ``Searches for the role of spin and polarization in gravity,''
  Rept.\ Prog.\ Phys.\  {\bf 73}, 056901 (2010)
  [arXiv:0912.5057 [gr-qc]].

\bibitem{LV-GR2}
M.L. Yan,
The renormalizability of general gravity theory with torsion and the spontaneous breaking of Lorentz group,
Commun.\ Theor.\ Phys.\ {\bf 2}, 1281 (1983).

\bibitem{Copenhagen}
H.B. Nielsen and M. Ninomiya,  $\beta$-function in a noncovariant
Yang-Mills theory, Nucl. Phys. B 141, 153 (1978);
\bibitem{Copenhagen2}
S. Chadha and H.B. Nielsen, Lorentz invariance as a low
energy phenomenon, Nucl. Phys. B 217, 125 (1983);
\bibitem{Copenhagen3}
H.B. Nielsen and I. Picek, The Redei-like model and testing
Lorentz invariance, Phys. Lett. 114B, 141 (1982);
\bibitem{Copenhagen4}
H.B. Nielsen and I. Picek, Lorentz non-invariance, Nucl. Phys. B 211, 269 (1983).

\bibitem{Ammosov2000}
V.~Ammosov and G.~Volkov, Can neutrinos probe extra
dimensions?, hep-ph/0008032.

\bibitem{Pas2005}
H. Pas, S. Pakvasa, T.J. Weiler,  Sterile-active neutrino
oscillations and shortcuts in the extra dimension, Phys.Rev. D {\bf 72}, 095017
(2005).



\bibitem{SME98}
D. Colladay, V. A. Kostelecky, Lorentz-violating extension
of the standard model, Phys. Rev. D {\bf58}, 116002 (1998).

\bibitem{KR08}
V. A. Kostelecky, N. Russell, Data tables for Lorentz and
CPT violation, Rev. Mod. Phys. {\bf83}, 11 (2011)
[arXiv:0801.0287].

\bibitem{Ma10}
L. Zhou, B.-Q. Ma, Lorentz invariance violation matrix from
a general principle, Mod. Phys. Lett. A {\bf25}, 2489 (2010)
[arXiv:1009.1331].

\bibitem{SMS3}
L. Zhou, B.-Q.~Ma, New theory of Lorentz violation from
a general principle, Chin.\ Phys.\ C (HEP \& NP) {\bf 35}, 987 (2011)
[arXiv:1109.6387].

\bibitem{Ma10graal}
L. Zhou, B.-Q. Ma, A Theoretical Diagnosis on Light Speed Anisotropy from GRAAL Experiment, Astropart.\ Phys.\ {\bf 36}, 37 (2012) [arXiv:1009.1675].

\bibitem{SMS-OPERA}
L. Zhou, B.-Q. Ma, Neutrino speed anomaly as signal of
Lorentz violation, Astropart.\ Phys.\ {\bf 44}, 24 (2013) [arXiv:1109.6097].


\bibitem{Ma:2011jj}
B.-Q.~Ma,
  ``The Phantom of the OPERA: Superluminal Neutrinos,''
 Mod.\ Phys.\ Lett.\  A {\bf 27}, 1230005 (2012) [arXiv:1111.7050].




\bibitem{BKLR03}
  R. Bluhm et al.,
  ``Probing Lorentz and CPT violation with space based experiments,''
  Phys. Rev. D {\bf68}, 125008 (2003).


\bibitem{Bocquet10}
J.-P. Bocquet et al., Limits on lightspeed
anisotropies from compton scattering of high-energy
electrons, Phys. Rev. Lett. {\bf104}, 241601 (2010).


\end{thebibliography}
\end{document}